# Chirp control of directional current in monolayer graphene by intense few-cycle laser

Erheng Wu[1,2,4], Qiang Zhan[1,2,4], Zhanshan Wang[1], Chaojin Zhang[3], and Chengpu Liu[2,*]

[1]*China MOE Key Laboratory of Advanced Micro-structured Materials, Institute of Precision Optical Engineering, School of Physics Science and Engineering, Tongji University, Shanghai 200092, China*
[2]*State Key Laboratory of High Field Laser Physics, Shanghai Institute of Optics and Fine Mechanics, Chinese Academy of Sciences, Shanghai 201800, China*
[3]*School of Physics and Electronic Engineering, Jiangsu Normal University, Xuzhou 221116, China*
[4]*University of Chinese Academy of Sciences, Beijing 100049, China*

*Corresponding author: **chpliu@siom.ac.cn**

**Abstract:** The residual current density in monolayer graphene driven by an intense few-cycle chirped laser pulse is investigated via numerical solution of the time-dependent Schrödinger equation. It is found that the residual current is sensitive to the initial chirp rate, and the defined asymmetry degree for current along the different polarization direction versus chirp rate follows a simple sinusoidal function. The underlying physical mechanism is the chirp-dependent Landau-Zener-Stückelberg interference. The chirp control of currents provides a novel convenient tool in the petaHertz switching of two-dimensional materials based optoelectronic devices on the sub-femtosecond timescale.

The interaction of a strong electric field with transparent crystalline solids is a long term research topic in the condensed matter physics, first introduced by Zener [1] and developed by Keldysh [2], and has recently drawn renewed attentions with the occurrence of atomically thin materials. For example, a strong few-cycle laser interacting with a two-dimensional narrow bandgap medium has makes the frontier of solid-state metrology pushed to PHz bandwidth and several hundred attosecond resolution [3] and a new class of phenomena in condensed matter optics. Here a strong optical field of 1-3 V/Å reversibly changes the solid within an optical cycle, and it is the *instantaneous* light-field beyond the *cycle-averaged* light intensity *dominating the microscopic dynamics of electrons in solids* [6].

The characteristic energies characterizing the light-matter interaction strength, such as Bloch frequency in dielectrics [4] or Rabi frequency in semiconductors [5] overtake the carrier wave frequency of the driving laser pulse, and thus the electron interband transition is greatly influenced by the electron intraband motion [6,7]. The microscopic dynamics of electrons is determined by the instantaneous light- field, indicating that the light-induced processes [8-17]are sensitive to the exact waveform of the driving laser. For example, a carrier-envelope-phase (CEP, or carrier phase) dependent current exhibiting a monotonic [*D. Sun, etal. Coherent control of ballistic photocurrents in multilayer epitaxial graphene using quantum interference, Nano. Lett. 10, 1293-1296 (2010)*] or non-monotonic behavior [7] with the increase of optical field strength has been demonstrated in zero-bandgap graphene. This non-monotonic behavior can even lead to a reversal of the current direction [8,18].

Beyond CEP, a multi-pulse superposition [18], pulse chirping [19] or other pulse shaping techniques [20] can also take the same role of a CEP. In this paper, an initially chirped few-cycle is thus used to investigated the non-monotonic behavior for current reversal in a monolayer graphene. It is confirmed that the residual current is really sensitive to the chirp rate, and the direction of residual current can be switched via the adjustment of chirp rate, which are the direct consequence of chirp-dependent Landau-Zener-Stückelberg (LZS) interference [21] on the sub-cycle timescale.

The light-field-driven electron dynamics in a monolayer graphene is investigated using a nearest-neighbor tight-binding model [8] under the driving of a few-cycle infrared chirped laser. The incident chirped field reads as $E_x = E_0 \exp\left(-(t/\tau_p)^2\right)\cos\left(\omega_0 t + 1/2\alpha t^2 + \varphi_0\right)$. which is linearly polarized along the $x$ direction. Here, $E_0$ refers field amplitude, corresponding to a peak intensity $I_0 = c\varepsilon_0 E_0^2/2$. $\omega$ is the driving field frequency and its corresponding central photon energy is 1.5eV when an infrared 800 nm laser used. In graphene, the strength of light-matter interaction can be characterized by Rabi frequency $\Omega_R$ with $\hbar\Omega_R = |v_F e\omega^{-1} E_0|$ [7]. To ensure that the light-graphene interaction reaches the light-field-driven regime, the laser relation $\Omega_R \geq \omega$ should be fulfilled ($E_0 > 1.8$ V nm$^{-1}$). Therefore, the driving field strength $E_0$ is selected as 2.4 V/nm. $\tau_p$ is pulse duration (here 5 fs is used, much shorter the characteristic electron scattering time of tens fs [3]. The pulse has a time-dependent carrier frequency $\omega = \omega_0 + \alpha t$. If $\alpha > 0$, the pulse is up-chirped, otherwise it is down-chirped. With pulse duration limits, the range of chirp rate from -0.07 to 0.07 (10$^{30}$ s$^{-2}$) is appropriate; the change of frequency induced by chirp $\alpha\tau_p$ approximately is 0.15$\omega_0$.

$\varphi_0$ is the carrier-envelop phase (CEP), i.e. absolute phase, which has widely confirmed its strong influence in the light-matter interaction processes, such as it influence on the control of asymmetric current generation in graphene[22]. Beyond the CEP's influence, here we only focus on the chirp rate $\alpha$ by setting $\varphi_0 = 0$ as zero.

The lattice structure of this monolayer graphene in the position space and reciprocal vector space are described in Figs. 1(a) and 1(b), respectively [23,24]. The positions of the nearest three atoms neighboring around a carbon atom at subsite A are $\boldsymbol{\delta_1}=a/2(-1,\sqrt{3})$, $\boldsymbol{\delta_2}=a/2(1,0)$ and $\boldsymbol{\delta_3}=a/2(-1,-\sqrt{3})$ with a lattice constant $a=0.246$ nm. In order to obtain the electronic band structure, the $p_z$ atomic orbitals are generally adopted to form a Bloch wave function. Thus, the corresponding field-free electron Hamiltonian is written as

$$H_0 = \begin{bmatrix} 0 & -\varepsilon_h f(\mathbf{k}_0) \\ -\varepsilon_h f^*(\mathbf{k}_0) & 0 \end{bmatrix}, \quad (1)$$

in which the hopping integral $\varepsilon_h = 3.0$ eV and $\mathbf{k_0}$ is the initial momentum of an electron. The form factor determining the symmetry property of a lattice structure takes the form of $f(\mathbf{k}_0) = \exp(i a k_{x0}/\sqrt{3}) + 2\exp(a k_{x0}/2\sqrt{3})\cos(a k_{y0}/2)$. The eigen-energies of conduction band $E_c(\mathbf{k_0}) = \varepsilon_h |f(\mathbf{k_0})|$ and valence band $E_c(\mathbf{k_0}) = -\varepsilon_h |f(\mathbf{k_0})|$ can be directly obtained by diagonalizing $H_0$ with the corresponding basis functions of

$$\Psi^v_{\mathbf{k}_0}(\mathbf{r}) = \frac{1}{\sqrt{2}}\begin{pmatrix} \exp(i\theta_{\mathbf{k}_0}/2) \\ \exp(-i\theta_{\mathbf{k}_0}/2) \end{pmatrix}, \Psi^c_{\mathbf{k}_0}(\mathbf{r}) = \frac{1}{\sqrt{2}}\begin{pmatrix} -\exp(i\theta_{\mathbf{k}_0}/2) \\ \exp(-i\theta_{\mathbf{k}_0}/2) \end{pmatrix}.$$

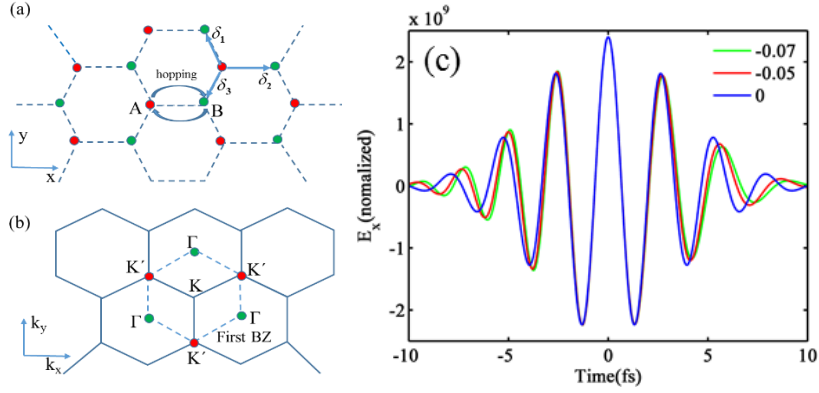

Fig. 1. Lattice structures of monolayer graphene in the (a) position space and (b) reciprocal vector space. A unit cell contains two carbon atoms at subsites A and B, and the three nearest neighbor positions around A are $\boldsymbol{\delta_1}=a/2(-1,\sqrt{3})$, $\boldsymbol{\delta_2}=a/2(0,1)$ and $\boldsymbol{\delta_3}=a/2(-1,-\sqrt{3})$. The first Brillouin zone (BZ) around an intervalley K are indicated by dashed lines. (c) Some representative laser waveforms with different initial chirp rates but a fixed peak strength of $E_0=2.4$V/nm.

Here, $\theta_{\mathbf{k}_0}$ is phase factor and defined as $\theta_{\mathbf{k}_0} = f(\mathbf{k_0})/|f(\mathbf{k_0})|$, the electron dynamics under ultra-short external field is coherent and can be described by the time-dependent Schrödinger equation (TDSE)

$$\frac{\partial}{\partial t}\Psi = -i\hbar^{-1}H\Psi, \tag{2}$$

with $H = H_0 - e\mathbf{E}(t)\cdot\mathbf{r}$. Generally speaking, under the driving of an intense short laser, the electron interband dynamics and the intraband dynamics are both important. The former is described by dipole coupling, while the latter follows the Bloch acceleration theorem:

$$\mathbf{k}(t) = \mathbf{k}_0 - e/\hbar \int_{-\infty}^{t} \mathbf{E}(t')dt'. \tag{3}$$

In Eq. (3), the electron intraband dynamics is naturally included, and thus it is convenient to solve TDSE on the basis of Huston states $\Phi_\beta[\mathbf{k}(t)]$ [25] with the general solution of $\Psi_{\mathbf{k}_0}(t) = \sum_{\beta=c,v} a_{\beta,\mathbf{k}_0}(t)\exp\left(\int_{-\infty}^{t} E_\beta[\mathbf{k}(t')]dt'\right)\Phi_\beta[\mathbf{k}(t)]$. This solution is substituted into Eq. (2) and then the temporal evolution of probability amplitude $a_{\mathbf{k}_0}^\beta(t)$ is obtained as

$$\dot{a}_{\mathbf{k}_0}^c(t) = -i\hbar^{-1}d_x(\mathbf{k}(t))E_x(t)\exp\left(\int_{-\infty}^{t} E_c[\mathbf{k}(t')] - E_v[\mathbf{k}(t')]dt'\right)a_{\mathbf{k}_0}^v(t),$$
$$\dot{a}_{\mathbf{k}_0}^v(t) = -i\hbar^{-1}d_x(\mathbf{k}(t))E_x(t)\exp\left(\int_{-\infty}^{t} E_v[\mathbf{k}(t')] - E_c[\mathbf{k}(t')]dt'\right)a_{\mathbf{k}_0}^c(t). \tag{4}$$

Here, the dipole matrix element $d_{x,\mathbf{k}_0} = \langle \Psi_{\mathbf{k}_0}^c | ex | \Psi_{\mathbf{k}_0}^v \rangle$. In the undoped graphene before laser excitation, all states in the valence band are completely occupied and those in the conduction band are empty. One can solve the above differential equations and obtain the dynamic occupation of conduction band as $\rho_{\mathbf{k}_0}^c(t) = |a_{\mathbf{k}_0}^c(t)|^2$. The chirp dependent residual current density is thus defined as $j_x(\alpha) = -2e/(2\pi)^2 \int_{BZ} \rho_{\mathbf{k}_0}^c(t_\infty)\mathbf{v}_F d\mathbf{k}$, [6,7], with $v_F$ indicating the Fermi velocity around K (K') point (~1 nm/fs) given by the slopes of the bands which are constant for zero-band monolayer graphene. Thus, from this formula of the residual current density, one can see that the magnitude of the current is determined by the integration of residual occupation of conduction band over BZ, indicating the more asymmetric the occupation, the larger the magnitude.

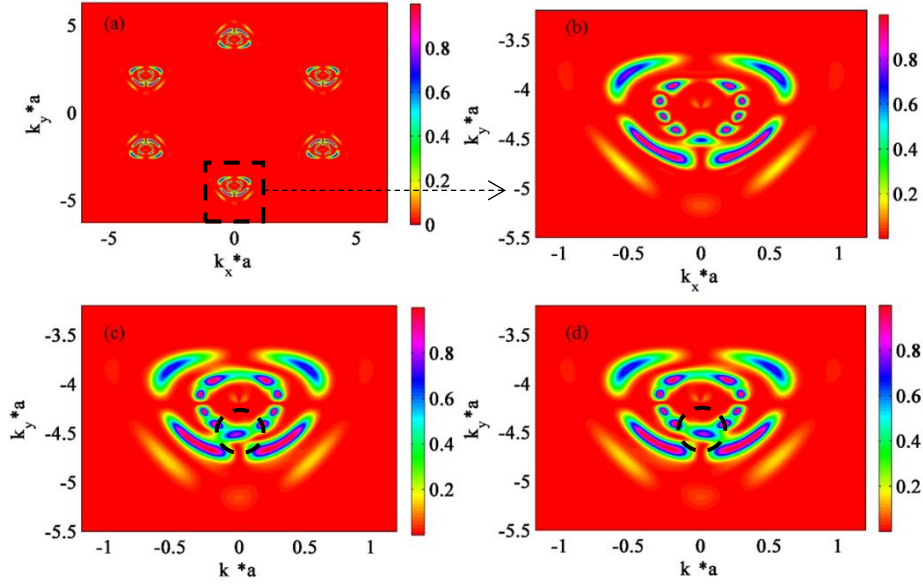

Fig. 2. Simulated residual conduction population distributions $\rho_{CB}$ for different initial chirp rates. (a) is the whole distribution in full Brillouin zone for $\alpha=0.00$ and the other are distributions in each intervalley K' for (b) $\alpha=0.00$, (c) $\alpha=-0.05$, and (d) $\alpha=0.05$

Figure 2 shows the residual occupations of electrons in the conduction band for different initial chirp rates. If chirp-free ($\alpha=0$), that is the incident laser pulse is transform limited, the population distribution is mirror symmetric along the laser polarization direction. Thus, the residual current after integration over momentum space is zero (Fig. 3). Beyond this, from inner to outside, clear distribution rings are exhibited (Fig. 2(a) which corresponding to one-photon absorption, two-photo absorption, and so on. When an initial chirp rate is introduced, such as $\alpha=\pm0.05$, the population distribution becomes asymmetric along the $k_x$ axis, manifested especially in the dashed circular areas in Figs. 2(c) and 2(d), where the residual population distributions present clear slopes, and the slope values for positive chirp ($\alpha=0.05$) and negative chirp ($\alpha=-0.05$) are with the opposite sign but the same magnitude (Fig. 3). This finally results in the residual current with the same magnitude but opposite sign.

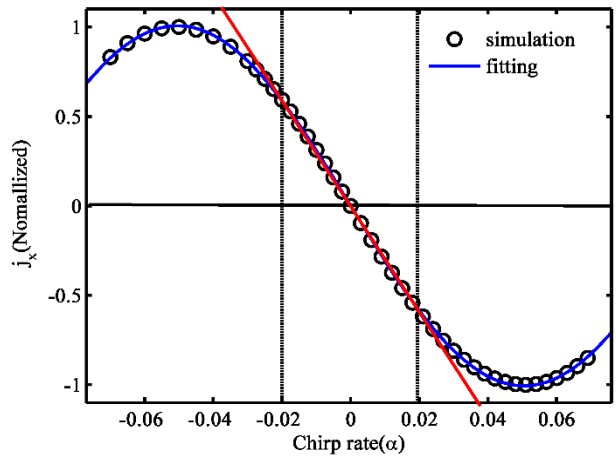

Fig. 3 Simulated residual current versus chirp rate, which is perfectly fitted by a sinusoidal function of sin ($\kappa\alpha$) ($\kappa$ is around 31.13 for the simulation parameters here)

With a small step change of chirp rate in the range of from -0.07 to 0.07, one can get a more clear relationship between the residual current density and chirp rate, as shown in Fig. 3. First, the residual current is very sensitive to chirp rate. More interesting, the whole chirp-dependence can be well fitted by a perfect sinusoidal function, sin ($\kappa\alpha$) with $\kappa$ is a suitable fitting factor. If chirp rate is small enough ($|\alpha| \leq \pm 0.02$), this chirp-dependence is simplified further as $\kappa\alpha$, indicating that the residual current is directly proportional to chirp rate (referring to the eye-guiding line in Fig. 3). There naturally occur two questions: One is why the residual current is chirp-dependent, and the other is why this chirp-dependence follows a sinusoidal function pattern. In the following, the underlying mechanism for them is disclosed in three steps. *Step* 1: as in the above demonstration, the residual conduction band population distribution is mirror asymmetric along the $k_x$ axis when an initial chirp rate is introduced [Fig. 2], and this asymmetry is chirp-rate-dependent. This point will be more clear if one making the chirp-dependent conduction band population difference, for example, between $\alpha = -0.05$ and $\alpha = 0.05$ (Fig. 4). In this condition, the population in $-k_x$ direction is larger than that in $+k_x$ direction, especially in regions around points M and P, leading to a negative current along $x$ direction.

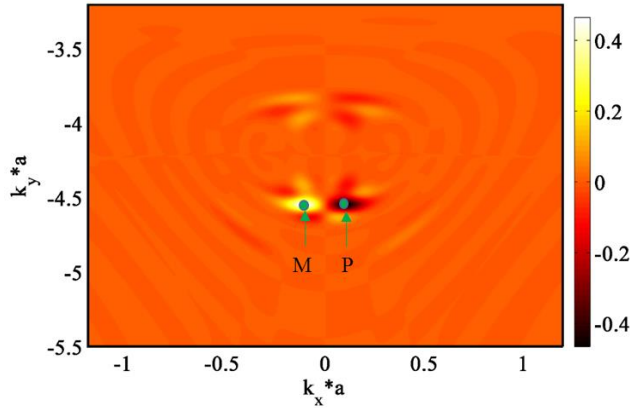

Fig. 4 Residual population difference of $\rho_{CB}(0.05) - \rho_{CB}(-0.05)$

*Step* 2: considering these two points P and M as examples, because they play the dominant contributions to the final asymmetric population distributions. Beyond this, the LZS interference is investigated to find the clues for the chirp-rate-dependence of residual current. In case of graphene, the transition probability can be estimated as $\exp(-\pi\Delta^2/4\hbar^2\Omega_R\omega)$ [7]. The electron dynamics driven by the external field follow the LZ formula, especially under condition of $\Delta \approx \sqrt{\Omega_R\omega}$, in which the electrons tend to pass the avoided crossings (i.e. bandgap minima), one part jumps non-adiabatically into the conduction band, while the rest still stays adiabatically in the valence band. As for a linearly polarized excitation, electrons can always repeatedly pass the avoided crossings within one optical period, leading the different excitation quantum pathways. These quantum pathways can interfere and thus the final electron population in the conduction would be sensitively dependent on the phase relationship among these pathways. To obtain the insight of LZS interference in light-field-driven regime, we need know the temporal evolution of the conduction band population. The case of LSZ interference is determined by two-phase terms. One is the transition phase (i.e. stokes phase) for a single LZ process between valence and conduction band, and the other is the propagation phase described as $\Delta\varphi_p = 1/\hbar \int_{t_1}^{t_2} E_c[\mathbf{k}(t')] - E_v[\mathbf{k}(t')] dt'$. Here $t_1$ and $t_2$ refer to the moments of two LZ transition events, and $E_c$ and $E_v$ represent the momentum-dependent energies of conduction and valence band states. Fig. 5 show the conduction band population and the propagation phase as a function of time for two different initial point M and P. For the trajectories starting from point M, the two transition events at approximately $t_1$ =0.0 fs and $t_2$ =1.2 fs in Fig. 5(d) (blue dashed lines). The propagation phase accumulation from $t_1$ to $t_2$ is π (Fig. 5(a)), and then plus the additional transition phase π from the LZ transition, the total phase accumulation would be 2π. Therefore, a constructive interference is induced. A larger population in conduction band occurs [Fig. 5(c), M point]. In contrast, the total phase accumulation is 3π from the start point P, resulting in a destructive interference and thus a smaller population in conduction band [Fig. 5(c), P point]. This kind of quantum phase difference for start points M and P finally lead to the occurrence of asymmetric population distribution in the strong field regime.

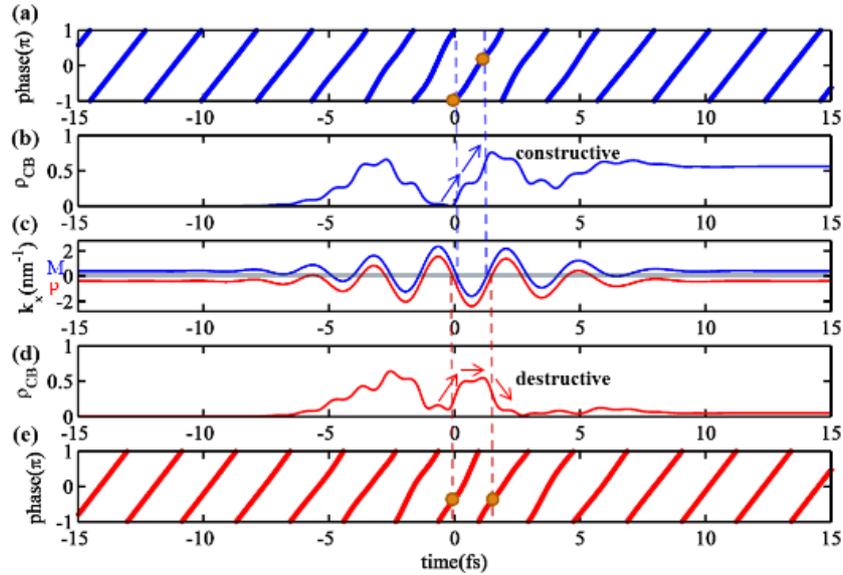

Fig. 5 Residual population $\rho_{CB}(0.05)$; The time evolution of propagation phases (a) (e) and conduction band populations (b) (d) for different initial wavenumber k at M and P points. (c) The two electron trajectories starting from points M and P in Fig. 4

*Step* 3: now let us answer why the relation between residual current density and chirp rate follows a sinusoidal function. As demonstrated above, the chirp-dependent residual asymmetric conduction-band population distribution originate in the pathways interference. In the following, we will explicitly write the chirp-dependent phase analytically. We define the magnitude of residual population in conduction band D in single k state [i.e. M or P point in Fig. 2(f)] in condition of two pathways $D = |\mathbf{A_1}| + |\mathbf{A_2}| + |\mathbf{A_1}||\mathbf{A_2}|\cos(\theta)$, with $|\mathbf{A_1}|$ and $|\mathbf{A_2}|$ indicating pathways probability amplitude between $(-\infty, t_1)$ and $(-\infty, t_2)$ and $\theta = \Delta\varphi_\mathrm{T} + \Delta\varphi_\mathrm{P}$. $D$ depends on their relative phase $\theta$. The transition phase $\Delta\varphi_\mathrm{T}$ is $\pi$, determined by the sign of electric field, which presents the avoided crossings[6], does not have difference with or without chirp rate. Therefore, we only focus on the propagation phase, which has the form $\Delta\varphi_\mathrm{p} = 2\varepsilon_h/\hbar \int_{t_1}^{t_2} |f[\mathbf{k}(t')]| dt'$ based on the dispersion relation of graphene. If the laser field *is x* polarized and the Dirac approximation is assumed,

$$\Delta\varphi_\mathrm{p} \approx 2\gamma v_F \int_{t_1}^{t_2} |k_x(t')| dt' = 2\gamma v_F \int_{t_1}^{t_2} \left|k_x(0) - \frac{e}{\hbar} A_x(t')\right| dt'. \qquad (5)$$

The integration term presents variation of electron momentum between transitions at time $t_1$ and $t_2$, and only has positive or negative condition, which introduces $\pi$ shift. This phase shift has not essential influence on the results. Thus, the absolute sign can be ignored. Since the vector potential $A_x(t)$ is the driving electric field under $\varphi_0 = 0$, takes the form $A_x(t) = E_0/\omega_0 f(t)\sin(\omega_0 t + 1/2 \alpha t^2)$, the above formula for propagation phase can be expanded as

$$\Delta\varphi_\mathrm{p} = 2\gamma v_F \int_{t_1}^{t_2} k_x(0) - \frac{eE_0}{\hbar\omega_0} f(t')\left[\sin(\omega_0 t')\cos(\alpha t'^2/2) + \cos(\omega_0 t')\sin(\alpha t'^2/2)\right] dt'.$$

If the chirp rate is small, one can further rewrite it as

$$\begin{aligned}\Delta\varphi_\mathrm{p} &= 2\gamma v_F k_x(0)\Delta t - 2\gamma v_F \frac{eE_0}{\hbar\omega_0} \int_{t_1}^{t_2} \sin(\omega_0 t') dt' \\ &\quad -\alpha\gamma v_F \frac{eE_0}{\hbar\omega_0} \int_{t_1}^{t_2} t'^2 \cos(\omega_0 t') dt' \qquad (6) \\ &= c_1 + \alpha c_2 \propto \alpha,\end{aligned}$$

with $c_1$ and $c_2$ are some constants. Thus, the population D is proportional to $\cos(\alpha)$. *As discussed above, the population in conduction band is sine relied on chirp rate and the residual current density is determined by the difference between critical* **k** *points. Based on the definition of residual current density, it follows sinusoidal function shape.*

Above all, via the three steps, the origin for the sinusoidal of chirp-rate-dependence of residual current is clarified.

**Conclusion**

We have numerically demonstrated that the electron intraband and interband motion in graphene can be controlled by adjusting the chirp rate of a few-cycle driving laser. In such a light-field-driven regime, the chirp-rate-dependent sub-optical-cycle Landau-Zener-Stückelberg interference causes the occurrence of an asymmetric residual ballistic electric current. This theoretical demonstration of ultrafast current control on the sub-femtosecond timescale will provide a meaningful guidance in the future experimental confirmation and development of solid-state petaHertz optoelectronic metrology.

**Acknowledgements**
The work is supported by National Natural Science Foundation of China (Grants No. 11374318 and No. 11674312)